\documentclass[10.5pt]{article}

\usepackage[utf8]{inputenc}
\usepackage[T1]{fontenc}
\usepackage{authblk}
\usepackage{abstract}
\usepackage{multicol}

\usepackage{caption}
\usepackage{amsmath,amssymb}
\usepackage{cite}
\usepackage[a4paper,left=1.8cm,right=1.8cm,top=1.8cm,bottom=1.8cm]{geometry}
\usepackage{hyperref}

\usepackage{graphicx}
\usepackage{dcolumn}
\usepackage{bm}

\usepackage{float}

\usepackage{epstopdf}%
\usepackage{amsfonts}%
\usepackage{amssymb}
\usepackage{xcolor}

\hypersetup{
    colorlinks=true,
    linkcolor=blue,
    citecolor=blue,
    urlcolor=blue
}

\title{\bf  A General Algorithm For Determining The Conductivity Zeros In Large Molecular Nanostructures: Applications To Rectangular Graphene Sheets}
\author[1]{M. Ni\c t\u a}
\author[1]{M. \c Tolea}
\author[2]{D. C. Marinescu}
\affil[1]{\small National Institute of Materials Physics, Atomistilor 405A,
Magurele 077125, Romania}
\affil[2]{\small Department of Physics, Clemson University, Clemson, SC 29634, USA}
\date{\today}

\begin{document}

\maketitle

\begin{onecolabstract}
\noindent
{\small
We propose an algorithm for determining the zeros of the electric conductivity in large molecular nanonstructures such as graphene sheets.
To this end, we employ the inverse graph method, whereby non-zeros of the Green's functions are represented graphically
by a segment connecting two atomic sites,
to visually signal the existence of a conductance zero as a line that is missing.
In rectangular graphene structures the
topological properties of the inverse graph determine the existence of two types of Green's function zeros
that correspond to absolute conductance cancellations
with distinct behavior in the presence of external disorder. We discuss these findings and their potential applications in some particular cases.}
\end{onecolabstract}

\vspace{1em}

\begin{multicols}{2}

\section{Introducere}

Electron transport in molecular structures is governed by the interference effects of the quantum states which determine the values of conductance.
Potential usage of such circuitry in real life application benefits from an apriori knowledge of the position of the conductance cancellations,
a desiderate that stimulated a significant body of research in the last decade
\cite{evers2020, tsuji2018, lambert2021, garner2020, gunasekaran2020, yangli2019, driscoll2021b, pan2022, valli2023, lambert2015,
kumar2024, fan2024, qu2022, ozlem2022, zhang2021}.
Moreover, the stability and control of these zeros under various external parameters are widely used
in the development of on-off devices such as transistors or switchers
\cite{tada2002, shuguang2018, cardamone2006, bones2021, tianming2023, alaa2023, chen2024, hector2024}.

Among the various methods that permit the detection of conductance zeros \cite{fowler2009,markussen2010,mayou2013,stuyver2015,nita2021},
we previously found that the graphic representation of the inverse matrix elements of the Hamiltonian $H$
as segments connecting the nodes of the molecular lattice can be readily used to predict the pairs of leads that do not support electric currents \cite{nita2022}.
This outcome is based on the fact that the inverse Hamiltonian matrix elements associated with any given two lattice points are
{equal to} the electron Green's function between the same points evaluated at zero energy.
The latter was shown to directly determine the electric conductance \cite{nita2021}.

{To pursue this approach one relies on} the existence of a discrete Hamiltonian,
such as the tight-binding model for graphene in solid state physics \cite{wallace1947} or the Hückel model for molecular systems \cite{tsuji2018}, which
arises from the localized spatial representation of atomic orbitals involved in the dynamics of conduction and valence electrons within the system.
{The inverse graph ${\cal G}_{inv}$ can take a simple form such as the complete bipartite graph for benzene or
a more complex structure as in the case of fulvene \cite{nita2022}.}
{It can be obtained by
various methods that a priori identify
the zeros of the Green's functions \cite{fowler2009, markussen2010, mayou2013, alexander2013, stuyver2015, tsuji2018, nita2021},
as well as by direct numerical evaluation of the inverse matrix.}

{In this paper we start by applying the inverse graph approach \cite{nita2022} to a series of graphene-like lattices.}
They are bipartite systems and
may differ from each other in their geometric characteristics. Although the dimension of the considered systems can be as large as possible,
here we establish some general principles that lead to their inverse graphs whose {structures are} shown to depend only on the graphene size in the armchair direction,
regardless of the size in the {zigzag} direction.



Then, within the Landauer-Büttiker formalism where the conductance zeros are associated with the zeros of the Green's function,
we classify the conductivity nulls based on their behavior in the presence of an external perturbation.
 {We find that the relevant distinguishing criterion is the
topological distance between the connecting nodes of the inverse graph, an integer value $d$ that is used
to define the order of a certain Green's function zero as $n=d-1$.}

In the case of rectangular graphene we demonstrate that two types of conductance zeros exist.
First-order zeros ${\bm G}_{ij}^{ (1)}$ appear between nodes separated by a distance $d(i,j)=2$.
We show that these points belong to the same sublattice of the bipartite system.
Second-order zeros, ${\bm G}_{ij}^{ (2)}$,
separated by distance $d(i,j)=3$ occur only between certain different partite points.
In the presence of disorder, the first order zeros are shown to be displaced to another energy value as the disorder strength increases, while those of the second order split, generating
 two dips in the conductance. Based on these findings, their usage as quantum on-off devices is discussed.

\section{The graphene lattice}

\begin{figure}[H]
\centering
\includegraphics[width=0.8\columnwidth]{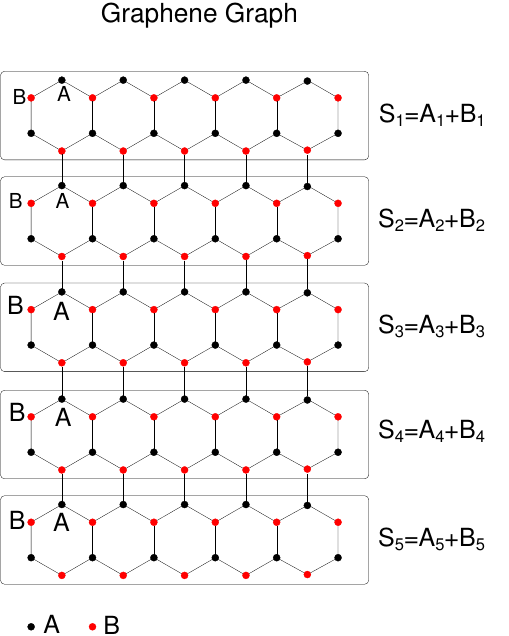}
\caption{\footnotesize
A picture of a rectangular graphene lattice
with $n_{zz}=5$ hexagons in the {{zigzag}} directions and $n_{ac}=5$ hexagons
in the arm-chair directions. It can be decomposed in $n_{ac}=5$ bipartite sublattices.
Every sublattice ${\bm S}_i$ with $i=1,\cdots, 5$ is a line of $n_{zz}$ hexagons and it is bipartite
with two subsets of points $A_i$ and $B_i$.
We select $A_i$ and $B_i$ sets such that the top {{zigzag}} edge belongs to $A_1$,
while the bottom {{zigzag}} edge is in $B_5$.
}
\label{grafena5b5z-1}
\end{figure}

We consider a class of graphene-type lattices, rectangular in shape,
with $n_{zz}$ hexagons in the {{zigzag}} directions and $n_{ac}$ hexagons
in the arm-chair directions. A lattice with $n_{zz}=5$ and $n_{ac}=5$ is shown in Fig.\,\ref{grafena5b5z-1}.
By varying $n_{zz}$ and $n_{ac}$,  different physical molecules are obtained, such as
benzene for $n_{zz}=1$ and $n_{ac}=1$,
 naphthalene for $n_{zz}=2$ and $n_{ac}=1$, biphenyl for $n_{zz}=1$ and $n_{ac}=2$
or perylene for {$n_{zz}=2$} and $n_{ac}=2$. All these systems have been extensively studied in the context of
single-electron molecular transport \cite{cardamone2006,tsuji2018,shuguang2018,evers2020,bones2021}.


The lattice in Fig.\,\ref{grafena5b5z-1} is bipartite, composed of two types of sites, $A$ (black) and $B$ (red).
 The Hamiltonian contains only hopping elements {between $(A,B)$ pairs of sites},
\begin{eqnarray}\label{hgr}
H=\sum_{i\in A, j\in B}  t_{ij}|i\rangle \langle j| +h.c.,
\end{eqnarray}
with $t_{ij}$ nonzero for all the graph lines depicted in Fig.~\ref{grafena5b5z-1}.
It is measured in the energy units $t$ whose typical value is $2.7$eV \cite{baer2002, valli2023}.
%

\begin{figure}[H]
\centering
\includegraphics[width=0.8\columnwidth]{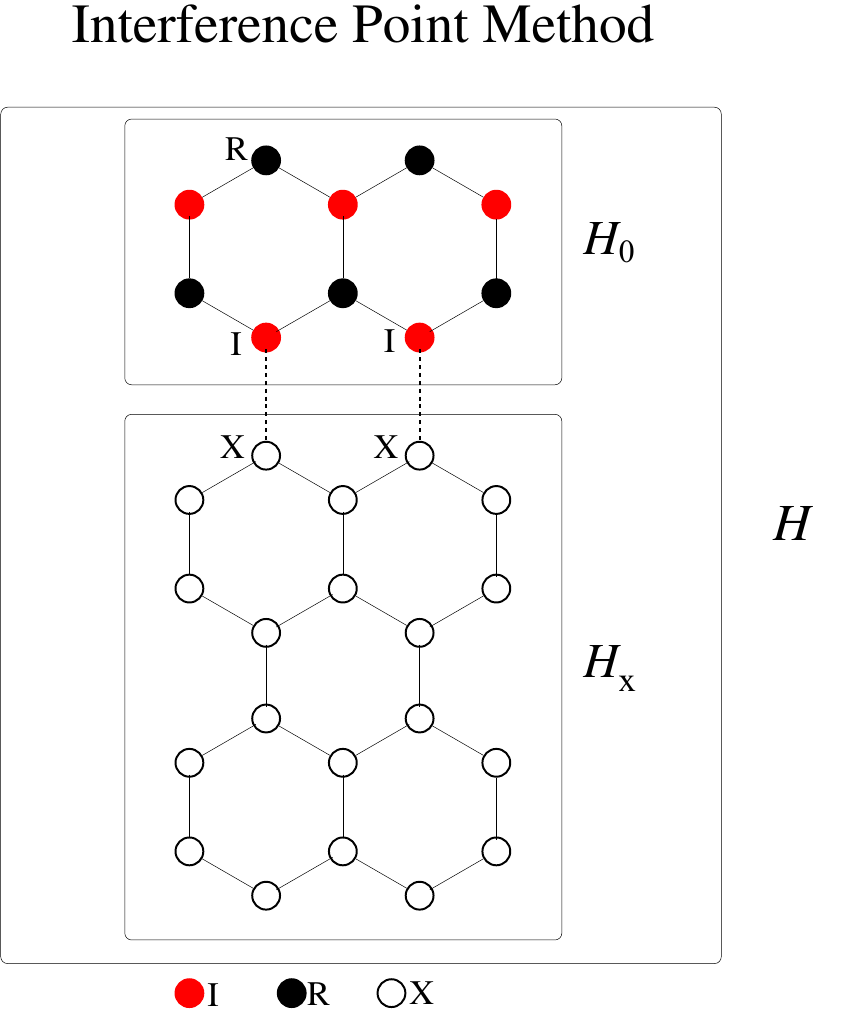}
\caption{\footnotesize
In the interference point algorithm, the original system described by $H$ is split in two subsystems.
The first one, $H_0$, contains the interference points $I$ that satisfy $G^0_{II}=0$ for any random pair,
and the points $R$. The second subsystem,
$H_X$, contains the points $X$, coupled only to points $I$. The DQI cancellations $G^0_{II}=0$ imply that $G_{IX}=0$ \cite{nita2021}.}
\label{mpi1}
\end{figure}

To determine the Green's function zeros in the graphene sheet we employ the interference point method, developed Ref.~\cite{nita2021}.
In this approach we identify a bi-partite, non-singular Hamiltonian $H_0$ based on a subset of points of the original lattice.
{As illustrated in Fig.\,\ref{mpi1},}
the lattice point set described by $H_0$ is the sum of two subsets: $I$, the interference set,
is identified based on the condition that there is a {destructive quantum interference (DQI)} between any two of its members,
\begin{eqnarray}\label{gii0}
G^0_{II}=0,
\end{eqnarray}
with $G^0$ the Green's function of $H_0$ at $E=0$; $R$ contains all the other points that are of no interest in the problem at hand.
In particular, when $I$ contains only $A$ or $B$ points, we write,
\begin{eqnarray}\label{gaag}
&& G_{AA}=0\;,\\
\label{gbbg}
&& G_{BB}=0\;.
\end{eqnarray}

The $(I,I)$ destructive interferences are robust to any lattice modification that involves only an $I$ site perturbation {\cite{nita2022}}.
To the lattice described by $H_0$, a set of $X$ points is added,
such that they perturb only the $I$ states, i.e. hopping exists only {between $X$} and $I$ points, but not between $X$ and $R$ points.

Thus, the original lattice described by the Hamiltonian $H$ is associated with {the set ${\cal M}$ whose structure is shown in Fig.\,\ref{mpi1}. This set is partitioned into three subsets, $\{I\}$, $\{R\}$, and $\{X\}$ such that $\{I\}\cup\{R\}$ are associated with the Hamiltonian $H_0$, as shown in Eq.~(\ref{mirx}),
\begin{eqnarray}\label{mirx}
{\cal M}= \overbrace{\underbrace{I+ R}_{H^0} +X}^{H}.
\end{eqnarray}
The Hamiltonian $H$ is consequently written as,
\begin{eqnarray}\label{hix}
H=H_0+H_X+V_{IX}+V_{XI},
\end{eqnarray}
with $V_{IX}$ and $V_{XI}$ designating hopping energies between $I$ and $X$ sites.}

This construction preserves the $(I,I)$ interferences which are now written for the Green's function of $H$,
\begin{eqnarray}
\label{gii}
G_{II}=0,
\end{eqnarray}
while    
additional DQI processes are obtained between pairs of points $(I,X)$,
\begin{eqnarray}
\label{gix}
G_{IX}=0,
\end{eqnarray}
indicating that Green's function zeros of the Hamiltonian $H$ are obtained from those of the Hamiltonian $H_0$.

{The above Green's function cancellations result from the Dyson expansions of $G_{II}$ and $G_{IX}$
considering $V$ in Eq.~(\ref{hix}) as a perturbation,
\begin{eqnarray}
G_{II}=&& G^0_{II}+G^0_{II} V_{IX}G_{XI} + G^0_{IX} V_{XI}G_{II} ,~\\
G_{IX}=&& G^0_{IX}+G^0_{II} V_{IX}G_{XX}+G^0_{IX} V_{XI}G_{IX}.~~~~
\end{eqnarray}
When $V=0$, the Green's functions $G^0_{II}$ and $G^0_{IX}$ are zero so the cancellations in Eqs.~(\ref{gii}) and (\ref{gix})
are obtained.}

We apply this method to graphene by identifying the lattice partitions that satisfy
Eq.\,(\ref{mirx}). To exemplify, we refer to the graphene sheet depicted in Fig.\ref{grafena5b5z-1} which has $n_{ac}=5$ bipartite sublattices, labeled by $\bm S_i$, with $i=1,\cdots, n_{ac}$. Each sublattice $\bm S_i$ contains partite sets, $A_i$ and $B_i$, serially coupled.

This structure allows a partition of the graphene sheet in
the three subsets of points $I$, $R$ and $X$ as in:
\begin{eqnarray}\label{irx2}
{\cal M}=\left(\underbrace{B_1}_{I} + \underbrace{A_1}_{R}   \right) + \left( \underbrace{S_2+S_3+S_4+S_5}_{X} \right).
\end{eqnarray}
In the absence of the set $X$, $S_1=A_1+B_1$ is the initial lattice described by the Hamiltonian $H_0$,
with $I=B_1$ the interference points and $R=A_1$ the rigid points.
Any propagation between the $I$ points leads to a DQI, i.e.
$
G^0_{B_1B_1}=0.
$
In the full system associated with the lattice ${\cal M}$ only the $I$ points are coupled
to the rest of the graph, $X=S_2+S_3+S_4+S_5$,
while none of the $R$ points have any coupling to $X$.
The partition shown in Eq.\,(\ref{irx2}) satisfies the criteria of the method outlined above, meaning that
DQIs exist between all pairs of points $I$ and $X$. Thus, with $I=B_1$ and $X=A_2,B_2,\cdots,A_5,B_5$, we find the following additional
$AB$ zeros,
\begin{eqnarray}\label{gb1a2}
G_{B_1A_2}=\cdots=G_{B_1A_5}=0.
\end{eqnarray}


The lattice partition in Eq.\,(\ref{irx2}) is not unique.
By increasing the size of the $I$ set,
new $AB$ zeros appear.
For instance, one can choose $S_1+S_2$ as the initial lattice, with $I=B_1+B_2$, $R=A_1+A_2$,
and $X=S_3+S_4+S_5$.
From Eq.\,(\ref{gix}) new $IX$ zeros emerge,
\begin{eqnarray}\label{gb2a3}
G_{B_2A_3}=G_{B_2A_4}=G_{B_2A_5}=0.
\end{eqnarray}


Finally,
the last choice for the $I$ set is
$I=B_1+\cdots+B_4$, while $R=A_1+\cdots+A_4$ and $X=S_5$.
Eq.\,(\ref{gix}) generates the last $AB$ zero
\begin{eqnarray}\label{gb4a5}
G_{B_4A_5}=0.
\end{eqnarray}

Eqs.~(\ref{gb1a2}), (\ref{gb2a3}), and (\ref{gb4a5})
show that the newly obtained zeros correspond to DQI processes between $B_i$ and $A_j$
lattice sites indexed by $i<j$.
This statement remains valid for any arbitrary rectangular graphene sheet
if one considers the following lattice decomposition,
\begin{eqnarray}\label{desc}
{\cal M}=\left( \underbrace{\sum_{i=1}^{n} B_i}_{I} + \underbrace{\sum_{i=1}^{n} A_i}_{R}  \right) +
         \left( \underbrace{\sum_{i=n+1}^{n_{ac}} S_{i}}_{X}                              \right),
\end{eqnarray}
for $n=1,\cdots,n_{ac}-1$.
In this case, the starting lattice described by $H_0$ is $S={S_1+\cdots+S_n}$.
{It is bipartite, with
$I = B$ and $R=A$}.
The $I$ points have the required properties for the application of the interference set method: there are the DQI process between all $I$ points, satisfying Eq.\,\ref{gii}; the composed lattice $I+R+X$ has the property
that only the $I$ points are perturbed by the $X$ addition, i.e. no $R$ point is coupled to any of $X$ sites.
The requirements of Eq.\,(\ref{mirx}) thus being fulfilled,
the $IX$ cancellations are obtained from Eq.(\ref{gix}).
With $I=B_1,\cdots,B_n$
and $X=A_{n+1},\cdots, A_{n_{ac}}$ the existence of new $AB$ zeros is proven.
These new zeros are valid for every value of $n$, so
 the general $AB$ zeros of the graphene lattice are obtained,
 \begin{eqnarray}\label{gabg}
 G_{B_iA_j}=0~\mbox{for}~i<j.
 \end{eqnarray}
The above Green's function cancellations are true for every rectangular graphene system as drawn in Fig.\,\ref{grafena5b5z-1},
regardless of the values $n_{ac}$ and $n_{zz}$. These are not, however, the typical zeros of bipartite lattices, such as $G_{AA} = 0$ or $G_{BB} = 0$,
as they are obtained from the properties of the interference points from Ref.\,\cite{nita2021}.

From Eqs.~(\ref{gaag}), (\ref{gbbg}), and (\ref{gabg}) which describe all the zeros obtained in the case of a graphene sheet, we conclude that
the only possible non-zero Green's functions at $E=0$ are,
 \begin{eqnarray}\label{gabf}
G_{B_i A_j}\ne 0~\mbox{for}~i \ge j.
\end{eqnarray}
In Table\,\ref{tabelgrafena}, we list all the AB graphene zeros from Eq.\,(\ref{gabg}) for $n_{ac}=5$.

\begin{table}[H]
\centering
\begin{tabular}{|c|c|c|c|c|c|}
\hline
$G_{AB}$ & $B_1$  & $B_2$  & $B_3$  &  $B_4$  &  $B_5$ \\
\hline
$A_1$    &&&&&  \\
\hline
$A_2$    &0&&&&  \\
\hline
$A_3$    &0&0&&&  \\
\hline
$A_4$    &0&0&0&&  \\
\hline
$A_5$    &0&0&0&0&  \\
\hline
\end{tabular}
\caption{\footnotesize The zeros of the $G_{AB}$ Green's function.}
\label{tabelgrafena}
\end{table}

\section{The Inverted Graph ${\cal G}_{inv}$.}

By using the  $AA$, $BB$, and $AB$ zeros from Eqs.\,(\ref{gaag}), (\ref{gbbg}), and (\ref{gabg}),
and the non-zero $AB$ functions from Eq.\,(\ref{gabf})
we are able to construct the Green's function graph.
This is the inverse graph of the graphene lattice and it has no line between $A$ points, no line between $B$ points,
and no line between $A,B$ pairs of points that satisfy the equality in Eq.\,(\ref{gabg}).

\begin{figure}[H]
\centering
\includegraphics[width=0.6\columnwidth]{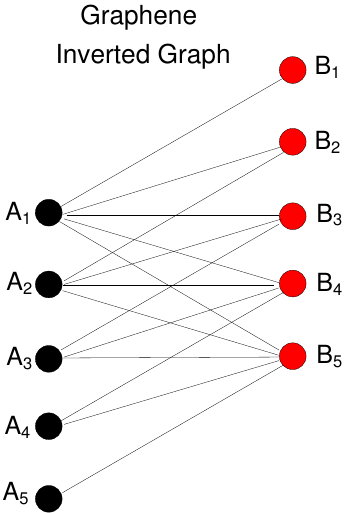}
\caption{\footnotesize
The general picture of the inverted graph for a rectangular graphene lattice.
The picture shows the non-zero elements of Green's functions at E=0.
The graphene lattice has $n_{ac}=5$.
}
\label{grafena5b5z-2}
\end{figure}

Since the inverse graph can be too complex/cluttered, depending on how many points are in the $A_i$ and $B_i$ sets,
in Fig.~\ref{grafena5b5z-2} we present a simplified version, where
the graph nodes are the sets of points $A_i$ and $B_i$ with $i=1,\cdots,n_{ac}$.
The graph lines are $(A_i, B_j)$ with $i\le j$ and they correspond to all non-zero Green's functions from
Eq.\,(\ref{gabf}).
For instance, the line $(A_1, B_4)$ from the simplified graph stands for what, in the original graph,
are a multitude of lines $(n,m)$ between the points $n\in A_1$ and $m\in B_4$.

A missing line in the graph means that the corresponding Green's function is zero.
For instance the graph contains no lines of the type $(A_i, B_j)$ with $i>j$ as they correspond to the Green's function cancellations from
Eq.\,(\ref{gabg}).

\section{Classification of conductivity zeros}

The existence of quantum tunneling between two lattice sites, $i,j$, implies that
the related Green's function is nonzero, $G_{ij}\ne 0$. On the graph this appears as a segment connecting
points $i$ and $j$. Thus, the graph distance between the two points is $d(i,j)=1$.
When the quantum propagation between points $i$ and $j$ leads to a DQI, $G_{ij}=0$.
This means that there is no direct connection between $i$ and $j$ and
the corresponding distance is given by the number of lines contained by the shortest path between $i$ and
$j$, leading to $d(i,j)>1$.

\begin{figure}[H]
\centering
\includegraphics[width=0.9\columnwidth]{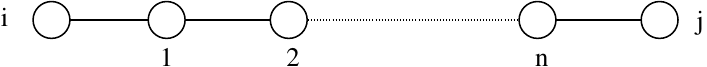}
\caption{\footnotesize
The shortest path between $i$ and $j$. The path belongs to the inverted graph ${\cal G}_{inv}$.
The graph distance $d(i, j)=n+1$. The DQI process $G_{ij}=0$ has the order $n=d-1$.
}
\label{gordn}
\end{figure}

A direct correlation can be established between the distance between two points and the behavior of the conductance zero in the presence of disorder.
We consider two points $(i,j)$ for which $G_{ij} = 0$ with $G$ the Green's function of the pristine Hamiltonian, $H$. This means that in the inverse graph ${\cal G}_{inv}$ there is no direct line $(i,j)$, but rather
the shortest path between $i$ and $j$ involves $n$ sites, the set ${\cal{P}}=\{1,2,\cdots, n\}$, as shown in Fig.~\ref{gordn}.

In the presence of a perturbation consisting of local disorder potentials,
\begin{eqnarray}\label{hprim-ap}
H'=H+\sum_{l\in{\cal M}} w_l|l\rangle \langle l |\;,
\end{eqnarray}
the Dyson expansion of the Green's function starts with the $n^{th}$ order term,
\begin{eqnarray}\label{gdeordo}
G'_{ij}= w_1w_2\cdots w_n {\cal C}_n + {\cal O}^{n+1}\;,
\end{eqnarray}
where ${\cal C}_n$ represents the products of the pristine Green's functions $G_{i1}G_{12}\cdots G_{nj}$ associated with the graph lines in Fig.\,\ref{gordn}.

By reduction to absurdum, we consider that there is a term of order $o<n$,
\begin{eqnarray}\label{gdeordo}
{G'}_{ij}^{(o)}= G_{i1}w_1G_{12}w_2\cdots w_{o-1}G_{o-1o} w_oG_{oj},
\end{eqnarray}
which contains the nonzero Green's functions $G_{i1},\cdots,G_{oj}$. This means
that there is a new path $(i,1,\cdots,o,j)$, with points $1,2,\cdots,o$ not necessarily included in ${\cal P}_n$,
of length $o+1$ smaller than the hypothesised shortest distance $n+1$ in Fig.\,\ref{gordn}.

In general, we can write that, for disorder strength $\epsilon$, for a distance $d(i,j) = n+1$, the Green's function is,
\begin{eqnarray}\label{gijen}
G'_{ij}=\epsilon^n {\cal C} + {\cal O}^{n+1}\;,
\end{eqnarray}
thus defining a zero of order $n$.

In the following considerations we use $d(i,j)$ as the principal parameter to investigate the
Green's function zeros. From the possible distance values in Fig.\ref{grafena5b5z-2}
one establishes
the existence of two types of Green's function zeros
in the graphene lattice.


\subsection{First order zeros}

We consider an $AA$ or $BB$ zero.
In bipartite lattices these are referred to as easy zeros \cite{tsuji2014,pedersen2014}, as they describe a DQI process that results from the quantum propagation between the same type of points.
Here, the two $A$ points belong to subsets $A_i$ and $A_j$, such that
$
G_{A_i A_j}=0$.
The properties of this zero are dictated by the graph distance $d(A_i, A_j)$ between the contact points involved.
Without loss of generality one chooses $i\le j$.


\begin{figure}[H]
\centering
\includegraphics[width=0.73\columnwidth]{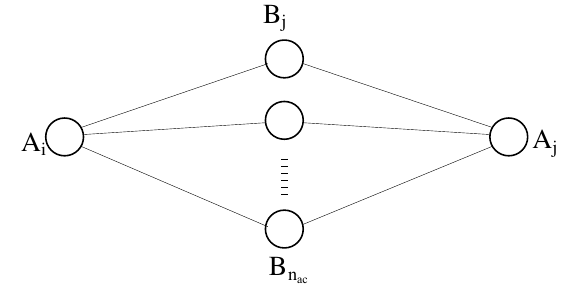}
\caption{\footnotesize
The shortest paths between two points of type $A_i$ and $A_j$ from the inverted graph.
The graph distance is $d(A_i,A_j)=2$
and the related DQI process $G_{A_iA_j}=0$ has order $n=d-1=1$. One chooses $i\le j$.
}
\label{gaa}
\end{figure}%

In the absence of a direct line between $A_i$ and $A_j$, we search for paths of length $2$.
The first order neighbors of the $A_i$ site form a set ${\cal N}^{A_i}$ that contains points $B_k$ with $k\ge i$ such that $G_{A_i B_k}\ne 0$, as it results from Eq.\,(\ref{gabf}),
\begin{eqnarray}\label{nai}
{\cal N}^{A_i}=\{B_i,B_{i+1},\cdots,B_j,B_{j+1},\cdots,B_{n_{ac}}   \}\;.
\end{eqnarray}
Similarly, the neighbours of $A_j$ form,
\begin{eqnarray}\label{naj}
{\cal N}^{A_j}=\{ B_j,B_{j+1},\cdots,B_{n_{ac}}  \}\;.
\end{eqnarray}
For $i\le j$, the common neighbours of $A_i$ and $A_j$
are given by ${\cal N}^{A_i} \cap {\cal N}^{A_j}$:
\begin{eqnarray}\label{msen}
{\cal M}_{sen}=\{B_j,B_{j+1},\cdots,B_{n_{ac}}   \}\;.
\end{eqnarray}
It means that the set formed out of graph paths $(A_i,B_k,A_j)$ with $k=j,\cdots,n_{ac}$ represents the shortest distance
between $A_i$ and $A_j$, as shown in Fig.\,\ref{gaa}. Thus, $
d(A_i,A_j)=2$. Correspondingly, the order of the $AA$ Green's function zero is $
n=d(A_i,A_j)-1 = 1$. This result is valid for all  $AA$ or $BB$ type of zeros.

\subsection{Second order zeros}

The second type of zeros encountered in the graphene lattice are of the $AB$ type, referred to as heavy zeros in the theory of bipartite lattices since
the end points belong to two different partitions of the lattice,
A and B respectively \cite{tsuji2014,pedersen2014, zhao2017}.
We consider such a zero given in Eq.\,(\ref{gabg}),  $G_{A_iB_j}=0$, with $i>j$.
To calculate its order we need to investigate the inverted graph distance between the two vertexes $A_i, B_j$.

\begin{figure}[H]
\centering
\includegraphics[width=0.92\columnwidth]{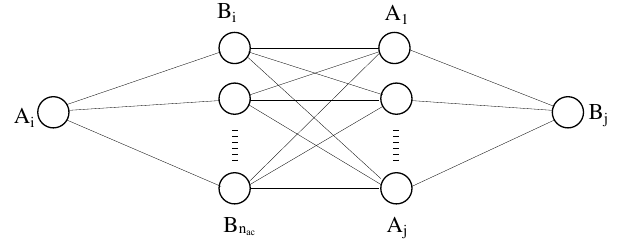}
\caption{\footnotesize
The shortest paths between $A_i$ and $B_j$ with $i>j$. The paths belong to the inverted graph.
The graph distance is $d(A_i, B_j)=3$.
The DQI process $G_{A_iB_j}=0$ has the order $n=d-1=2$.
}
\label{gab}
\end{figure}

First, there is no direct line $(A_i,B_j)$ due to Green's function cancellations.
The first-order neighbours of $A_i$ and $B_j$ are pictured in Fig.\,(\ref{gab}).
The set ${\cal N}^{A_i}$ has been already given in Eq.\,(\ref{nai}), while the first-order neighbors of $B_j$ form the set,
\begin{eqnarray}\label{nbj}
{\cal N}^{B_j}=\{ A_1,\cdots,A_{j}  \}.
\end{eqnarray}
Since ${\cal N}^{A_j}\cap {\cal N}^{B_j}=\emptyset$ there is no path of length $2$ from $A_i$ to $B_j$.
It is possible, however, to construct a path of length $3$.
From Eq.\,(\ref{gabf}), for $i>j$, it follows that we have a graph segment between all pair of points $({\cal N}^{A_i},{\cal N}^{B_j})$.
The corresponding lines are drawn in Fig.\,\ref{gab} which shows that the shortest paths between $A_i,B_j$ {have the} length
$
d(A_i, B_j)=3$.
Accordingly, the $AB$ zeros have order $n=d-1=2$.

\section{The effect of disorder}

In this section we analyze the behavior of the conductance zeros in the presence of disorder. A zero of the Green's function, associated with destructive quantum interference,
leads to the cancellation of electrical conductance in the system connected to transport wires \cite{tsuji2018,nita2021}.
These zeros can be immune to an external perturbation that affects only certain sites of the lattice \cite{shuguang2018, junyang2018, nita2022},
or they can be modified, with different scenarios already proposed in the literature.
The shifting of the zeros to other energy values or their splitting can occur in the presence of local impurities or off-diagonal energies
\cite{tsuji2014, pedersen2014, garner2016, sara2018, tsuji2019, valli2023}.

We consider an external perturbation that induces on-site energies $w_l$ on the lattice sites $l\in {\cal M}$,
\begin{eqnarray}\label{hprim}
H'=H+\sum_{l\in{\cal M}} w_l|l\rangle \langle l |\;,
\end{eqnarray}
with $w_l$ randomly distributed in the interval of width $W$,
\begin{eqnarray}
w_l \in \left(-W/2,W/2 \right).
\end{eqnarray}
$W$ is called the Anderson disorder strength.

For a small energy $E\simeq 0$,
the lowest order terms of $G'(E)=\frac{1}{E-H'}$
are calculated by using the Dyson expansion formula (see Eq.\,(\ref{gdeordo})).
The first order zero $G'_{A_i A_j}$ contains only the on-site energies applied on the $B_k$ lattice
sites that belong to the shortest paths $(A_i, B_k, A_j)$ from Fig.\,\ref{gaa}, leading to
\begin{eqnarray}\label{g1new}
G'_{A_i A_j}(E)=\sum_{k\in {\cal M}_{sen}} (w_k-E) {\cal C}_k + {\cal O}^2\;,
\end{eqnarray}
with ${\cal C}_k$ representing the pristine Green's function products $G_{A_i k}G_{k A_j}$
and $k$ a label for the common neighbours of $A_i,A_j$  defined in Eq.\,(\ref{msen}).

Therefore, in the presence of disorder, at $E=0$ the Green's function has a finite value, and the DQI process is shifted to an energy given by the solution of
$G'_{A_i A_j}(E)=0$,
a first order equation in $E$.

\begin{figure}[H]
\centering
\includegraphics[width=0.6\columnwidth,angle=-90]{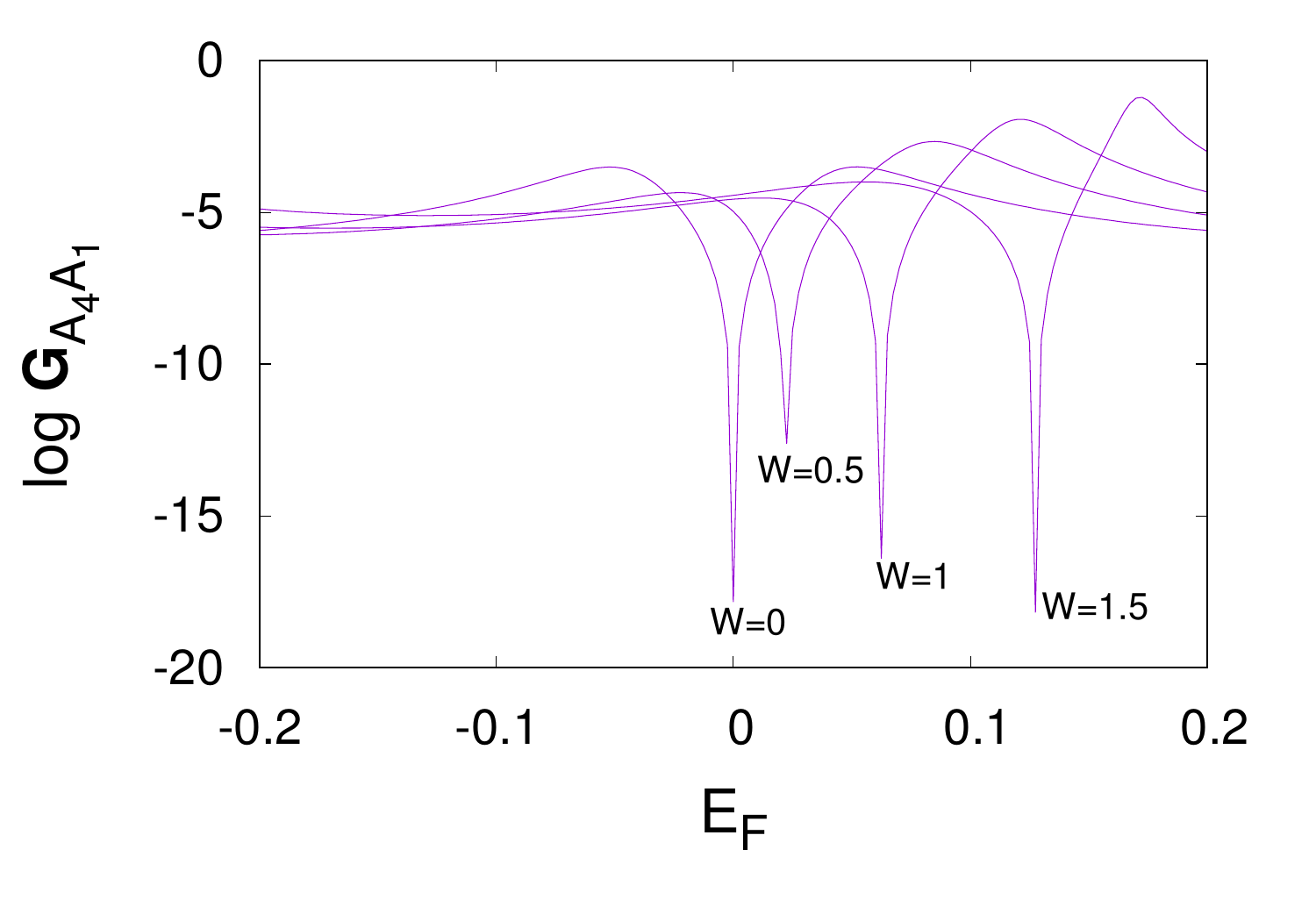}
\caption{\footnotesize
The displacement as a function of energy of a first order zero ${\bm G}_{AA}$ in the presence of disorder.
For low disorder, the shift to the new energy of the DQI process $E'$ increases with the disorder strength.}
\label{taa}
\end{figure}

This behavior is reproduced by the corresponding conductances ${\bm G}_{A_iA_j}$, a result reflected
by the numerical simulation in Fig.\,\ref{taa} in which we considered
a graphene lattice with $n_{ac}=4$ and $n_{zz}=3$ with
two external transport leads attached to sites from the $A_4$ and $A_1$ sets.
By using the effective Hamiltonian method from \cite{ostahie2021}
that considers the contact sites of the transport leads,
the conductance is calculated from the Landauer-B\"uttiker ansatz \cite{landauer1957,buttiker1986,horia2005}.
The Fermy energy is varying in the interval $E_F\in(-0.2,0.2)$.
{The small increase of the conductance maximum with disorder corroborates with the delocalization effects
affecting the edge states as discussed in Ref.~\cite{nita2018}.}

As shown, as disorder increases, the conductance zeros from $E=0$ in the clean limit ($W=0$) migrate to different energies $E'$ when
$W=0.5, 1$, and $1.5$.
As expected, in the low perturbation limit, the absolute value of $E'$ increases with disorder strength $W$.


The Dyson expansion of the second order Green's functions $G'_{A_iB_j}$ contains no first order terms, as we have no
path of length $2$ between $A_i$ and $B_j$.
In this case, for small energies values one obtains,
%
{\footnotesize
\begin{eqnarray}\label{g2new}
G'_{A_i B_j}(E)=\sum_{ k,l \in {\cal N}^{A_i}, {\cal N}^{B_j}}  (w_k-E) (w_l-E) {\cal C}_{kl}
+{\cal O}^3.
\end{eqnarray}}
The labels $k,l$ index the sites on the shortest paths from Fig.\,\ref{gab}, while
the constant ${\cal C}_{kl}$ contains the products of pristine Green's functions and can be derived from the
Dyson expansion formula (see for instance Eq.\,(\ref{gdeordo})).

\begin{figure}[H]
\centering
\rotatebox{-90}{\includegraphics[width=0.6\columnwidth]{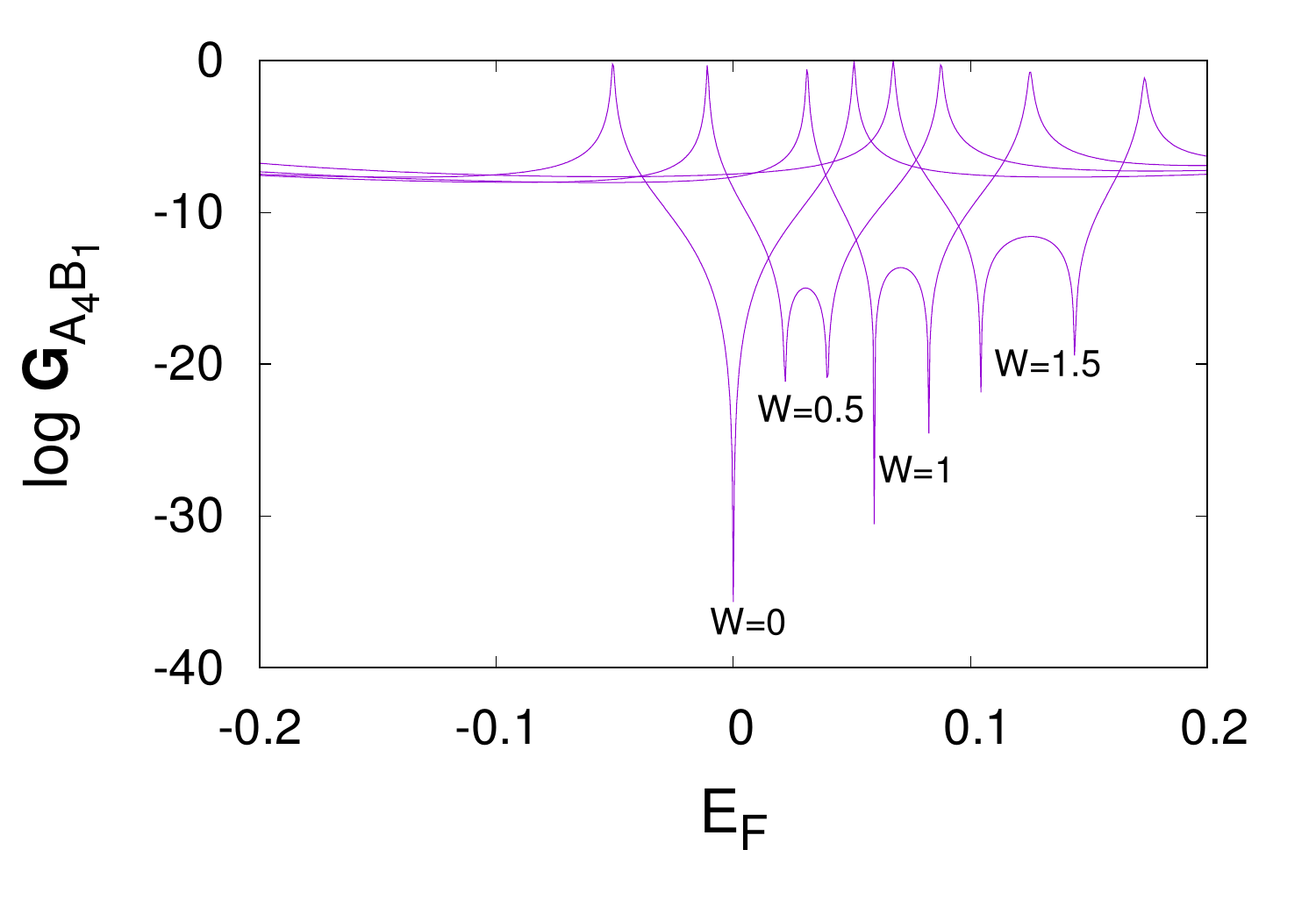}}
\caption{\footnotesize
The evolution of the second order zero ${\bm G}_{AB}$ with the disorder strength $W$
for a graphene lattice.
The dip of the zero conductance at $E_F=0$ is split, showing two new zeros at $E_1$ and $E_2$
in the presence of disorder.
The splitting energy $|E_1-E_2|$ increases with disorder strength $W$.
}
\label{tab}
\end{figure}

The equation $G_{A_i B_j}(E)=0$
 is quadratic in $E$, its two solutions $E_1, E_2$ indicating that the second order DQI process is split and two conductance zeros arise at Fermy energies $E_F=E_1$ and $E_F=E_2$. A numerical example is shown in Fig.\,\ref{tab}
for the same lattice as in Fig.\,\ref{taa}.
The conductance is calculated for contact sites from the sets $A_4$ and $B_1$.
While a single conductance dip at $E_F=0$ exists for a clean system, increasing the disorder magnitude leads to the splitting of the
original DQI in two conductance dips at $E_1$ and $E_2$ for low disorder.
The energy splitting $|E_1-E_2|$ increases with the disorder strength $W$.
{Other details of the numerical calculations related to Figs.~\ref{taa} and \ref{tab} are given in the Appendix.}


 \section{Applications}

The Dyson expansions in Eqs.~(\ref{g1new}) and (\ref{g2new})
show us how the related conductances
can be used in devices of the on-off type.
For $E=0$, from the first order zero one obtains
\begin{eqnarray}\label{gdewk}
G'_{A_i A_j}=\sum_{k\in {\cal M}_{sen}} w_k {\cal C}_k + {\cal O}^2\;.
\end{eqnarray}
This suggests that only one single site perturbation, applied on $k$ from the shortest path from $A_i$ and $A_j$,
can turn on the conductance zero.

For a second order zero
under the external perturbations, the Dyson expansion in Eq.\,(\ref{g2new}) at $E=0$ becomes
\begin{eqnarray}\label{gaibj3}
{G'}_{A_iB_j}= \sum _{ k,l \in {\cal N}^{A_i}, {\cal N}^{B_j}} w_k w_l  {\cal C}_{kl} + {\cal O}^3\;.
\end{eqnarray}
To use a second order zero to design an on-off switching device, two simultaneous perturbation have
to be applied on two sites
from the shortest path of Fig.\,\ref{gab}.

The related conductances can be also used to design logical gates. OR gates are obtained by using {the first order} zero
from Eq.\,(\ref{gdewk})
and AND gates by using {the second order} zero from Eq.\,(\ref{gaibj3}).
All such quantum devices that can be constructed from the {two types} of zeros of the graphene lattices
are listed in Table \ref{lggrafena}.

\begin{table}[H]
\centering
\resizebox{\columnwidth}{!}{  
\begin{tabular}{|c|c|c|c|}
\hline
Output       & Logical        & Location of   & Location of    \\
                  & Gate   & Input 1        & Input 2        \\
\hline
${\bm G}_{A_nA_m}^{(1)}$    & OR & $B_{\max(n,m)},\cdots, B_{n_{ac}}$ & $B_{\max(n,m)},\cdots, B_{n_{ac}}$ \\
$\forall n,m$  &&&   \\
\hline
${\bm G}_{A_nB_m}^{(2)}$   & AND & $B_n, B_{n+1},\cdots, B_{n_{ac}}$  &  $ A_1, A_2,\cdots, A_m$   \\
$n>m$                      &&&  \\
\hline
\end{tabular}
}
\caption{\footnotesize
Logic gates that can be constructed with the first and second order zeros of the rectangular graphene.
The output conductances are measured at Fermi energy $E_F=0$ and the two input parameters are the on-site perturbations
written in the third and fourth columns.}
\label{lggrafena}
\end{table}

Interestingly enough, the types of zeros that are obtained
do not depend on the number of hexagons in the zig-zag direction, $n_{zz}$,
but only on the number of hexagons in the armchair direction, $n_{ac}$.
Based on these considerations, for $n_{ac}=1$, one obtains a series of molecules,
benzene for $n_{zz}=1$, naphthalene for $n_{zz}=1$, anthracene for $n_{zz}=3$, and so on ...,
all of them having {only first order zeros}.
Below we illustrate our theory for $n_{ac}\ge 2$, when both types of zeros are obtained.

\begin{figure}[H]
\centering
\includegraphics[width=0.9\columnwidth]{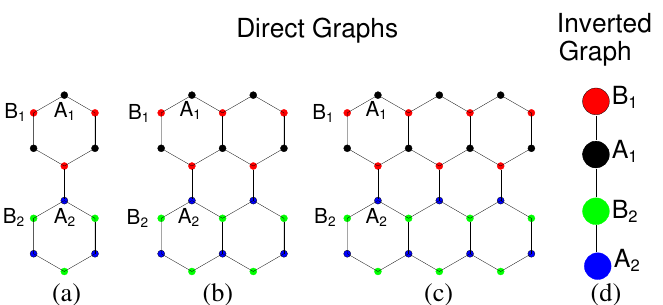}
\caption{\footnotesize
The biphenyl series.
The first three molecules from the series of lattices with $n_{ac}=2$:
the biphenyl for $n_{zz}=1$ in (a), the perylene for  $n_{zz}=2$ in (b) and bisanthene for $n_{zz}=3$ in (c).
All the molecules with $n_{ac}=2$, for all values $n_{zz}$, have the same structure of the inverse graph that is pictured in (d).
}
\label{bifenil}
\end{figure}

For $n_{ac}=2$, several different molecules are realized:
biphenyl for $n_{zz}=1$,
perylene for $n_{zz}=2$,
or bisanthene for $n_{zz}=3$, pictured in
Fig.\,\ref{bifenil}(a), (b) and (c).

All these molecules have two subsets in each partite set,
$A=A_1+A_2$ and $B=B_1+B_2$,
in agreement with the decomposition in Fig. \ref{grafena5b5z-1}.
Regardless of the value of $n_{zz}$ they all have the same inverse graph structure that is shown in Fig. \ref{bifenil}(d).

These molecules have three types of first order zeros, $G_{A_1A_2}, G_{A_1A_1}$ and $G_{A_2A_2}$.
The conductance zero associated with  $G_{A_1A_2}$ corresponds to the shortest path between $A_1, A_2$,
with ${\cal P}=(A_1, B_2, A_2)$. Consequently,
 a single site perturbation $w_k$ applied in $B_2$ is enough to displace it.

The molecules in the biphenyl series exhibit second order zeros such as $G_{A_2B_1}$, which is associated with a
path set ${\cal P}=\{B_1,A_1,B_2,A_2\}$. In this case,
two on-site energies $w_k, w_l$ have to be applied in the intermediate points $ A_1$ and $B_2$ for
the related conductance to become nonzero. A summary of these results is presented in Table 3.

\begin{table}[H]
\centering
\resizebox{0.9\columnwidth}{!}{
\begin{tabular}{|c|c|c|}
\hline
Conductance  & Type         & Who can destroy           \\
            &               &  this zero              \\
\hline
${\bm G}_{A_1A_1}$   & 1 & $w_k$, $k\in B_1+B_2$    \\
\hline
${\bm G}_{A_1A_2}$   & 1 & $w_k$, $k\in B_2$    \\
\hline
${\bm G}_{A_2 A_2}$   & 1 & $w_k$,  $k \in B_2$   \\
\hline
${\bm G}_{A_2B_1}$   & 2 & $w_kw_l$, $k\in A_1$,  $l\in B_2$   \\
\hline
\end{tabular}}
\caption{\footnotesize {The first and second order zeros} of all the molecules in the biphenyl series.
{The first order zeros} are destroyed by a one-site perturbation $w_k$ which shifts them from $E=0$.
{The second order zeros} are destroyed by two single-site energies simultaneously applied which {split} the original antiresonance dip.
\label{lgbifenil}}
\end{table}

{The algorithm outlined above enables the identification of a specific site $B$ that can modify an $AA$ zero in graphene in any general situation, regardless of the exact nature of the applied perturbation. This is also the case of inelastic collisions in the low order expansion (LOE) of transport.
For instance, the only inelastic collisions that can perturb $G_{A_2A_3}=0$ in Fig.\,3,
are
associated with the common neighbours of $A_2$ and $A_3$, which are $B_3$, $B_4$ and $B_5$ (see {the first} order terms in Eq.\,27).
In contrast, {the second-order} zeros are immune to inelastic collisions in LOE. For example, the $G_{A_4B_2}$ zero from Fig.~3,
remains invariant under any inelastic collisions
as the sites $A_4$ and $B_2$ have no common neighbours.
These results underscore that in general, the first-order zeros (as $G_{AA}$ or $G_{BB}$)
are affected by inelastic collisions, while the second-order zeroes are not.
These findings are consistent with the results presented in Ref.\,\cite{tsuji2018b} for $G_{AA}$ and $G_{AB}$ zeroes in linear polyene,
a system that is topological identical with graphene
(a linear polyene with 10 atoms and a graphene with $n_{ac}=5$ have the same type of inverse graph as in Fig.\,3).}

From these considerations, we conclude that on-off devices that rely on {the second-order} conductance as their output - such as $G_{AB}$ zeros in graphene-
are robust under hostile environmental conditions.

\section{Conclusions}

In this article, we generalized the inverse graph method
to calculate the structure of the inverse graph ${\cal G}_{inv}$ for systems of the rectangular graphene type.
For a given armchair length $n_{ac}$, the structure of ${\cal G}_{inv}$ has the same pattern for any
value of the zig-zag length $n_{zz}=1,\cdots, \infty$.

The conductance zeros are ranked according to the topological distance between the contact
sites, measured on the inverse graph.
In rectangular graphene there are two types of conductance zeros that
exhibiting distinct properties under external perturbations.
First, we find that
zeros of the first order ${\bm G}_{ij}^{(1)}=0$ correspond to points $i,j$ with the distance on the graph equal to 2.
In the presence of the local disorder potential, they are modified by the impurities located
on neighboring sites of both $i$ and $j$ points and
their energy position is shifted with increasing disorder strength.

Then, the
zeros of the second order ${\bm G}_{ij}^{(2)}=0$ appear between those points $i,j \in {\cal G}_{inv}$
having a graph distance equal to 3.
They are invariant to the application of any single site perturbation, and can be modified by
at least two impurities located in certain specific points $k,l$ that belong to one of the shortest pathes between $i$ and $j$.
With increasing disorder strength this zero is split into
two deep antiresonances.

Potential applications to on-off quantum devices that can be constructed with these conductances have been discussed.

~

\newpage

\appendix

\section{{The General formula of quantum conductance}}


{The widely adopted framework for calculating the electric conductance in coherent quantum devices is the
Landauer-B{\"u}ttiker formalism \cite{diventra2008, landauer1957, buttiker1986}.
In this approach, two terminals (transport leads, electrodes or external atoms chains) are attached
to discrete sites $i$ and $j$ of a quantum system as presented in \cite{tada2002, pedersen2014, markussen2010, shuguang2018}, while some
 experimental methods are addressed in \cite{evers2020}.}

{An electron with energy $E$ enters the system through the terminal attached to site $j$ and scatters out through the terminal attached to site $i$.
The Landauer conductance is expressed in terms of the tunneling amplitude between the two terminals \cite{evers2020,tsuji2018b}:
\begin{eqnarray}\label{land}
{\bm G}_{ij}(E) = \frac{e^2}{h} {\mbox{Tr}} [ \Gamma_i(E) G^R(E) \Gamma_j(E) G^A(E) ].
\end{eqnarray}
$\Gamma_i(E), \Gamma_j(E)$ are the scattering rates due to the couplings {$\tau_i$ and $\tau_j$} between the quantum system and external terminals.
$G^{R,A}$ are retarded and advanced Green's functions that can be calculated by using the
effective Hamiltonian
{$H^{eff}=H+\tau_i e^{-ik}|i\rangle\langle i|+  \tau_j e^{-ik} |j\rangle\langle j|$}
 \cite{ostahie2021}, {with $\tau_i=\tau_j={\tau_c^2}/{\tau_l}$ for a symmetric coupling. Here,
 $\tau_l$ is the hopping energy on the leads and $\tau_c$ is the coupling energy between leads and sample.
The electron energy is $E=2\tau_l \cos k$.}

{
The tunneling amplitude for an energy $E$ is numerically calculated from the effective Green function \cite{ostahie2021, nita2021} as,
\begin{eqnarray}\label{tun}
{\bf t}=2i \sqrt{\tau_i\tau_j} G_{ij}^{eff}(E)\sin k,
\end{eqnarray}
with $G^{eff}(E)=(E-H^{eff})^{-1}$.
Its square modulus generates the numerical conductance in $e^2/h$ units.}

{For the contact points indicated in Fig.\,\ref{fire} we compute two types of conductance exhibiting different behavior in the presence of disorder:
the first-order conductance ${\bf G}_{AA}$ (Fig.\,\ref{taa}), which shifts as disorder increases,
and the second-order conductance ${\bf G}_{AB}$ (Fig.\,\ref{tab}), which exhibits a splitting behavior.
The parameters for the numerical calculation are $\tau_l=2$ and $\tau_c=1$, measured in energy units $t$.}


\begin{figure}[H]
\centering
\includegraphics[width=0.9\columnwidth]{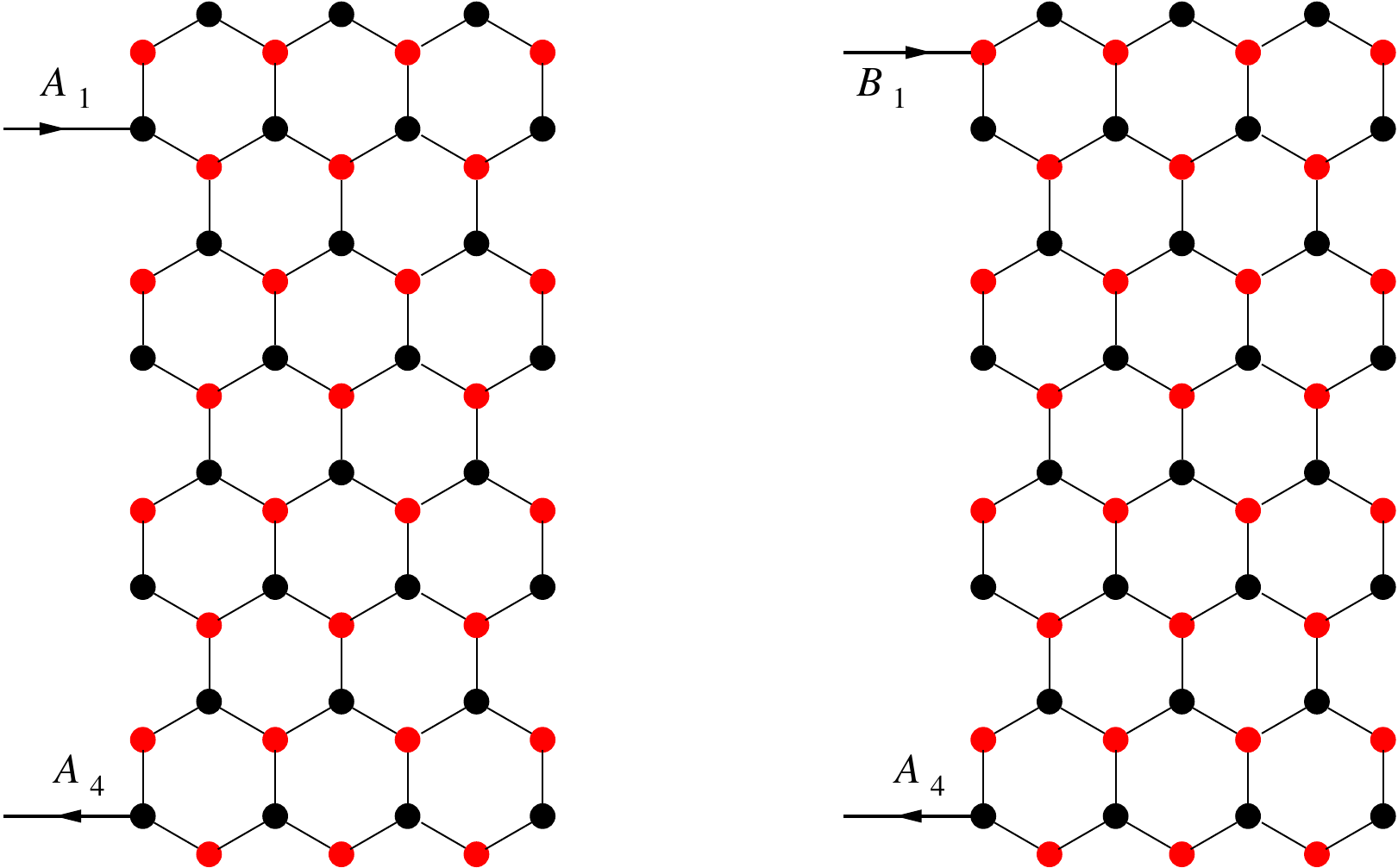}
\caption{\footnotesize
{
The transport lead position for the numerical calculation of the first order conductance ${\bf G}_{A_4A_1}$ and
of the second order conductance ${\bf G}_{A_4B_1}$.}}

\label{fire}
\end{figure}

{When $E=0$, corresponding to the wave number $k=\pi/2$, the tunneling amplitude in Eq.~(\ref{tun})
is calculated by using the Dyson expansion of the effective Green functions versus the coupling potentials in $H^{eff}$.
The conductance ${\bm G}=\frac{e^2}{h}|{\bf t}|^2$} is obtained \cite{tsuji2018, nita2021}:
{\footnotesize
\begin{eqnarray}\label{ap1}
{\bm G}_{ij} = \frac{\frac{e^2}{h} 4\tau_i\tau_j |G_{ij}|^2}
{|1 +i \tau_j G_{jj} +i \tau_i G_{ii} + \tau_i\tau_j |G_{ij}|^2 - \tau_i\tau_j G_{ii} G_{jj}|^2}.
\end{eqnarray}}
Here, $G_{ij}$, $G_{ii}$, and $G_{jj}$ are matrix elements of the Green's function, $G(E) = \frac{1}{E-H}$, evaluated at $E=0$.}

{Equation~(\ref{ap1}) shows that destructive quantum interference (DQI) in the tunneling between lattice sites
$i$ and $j$ occurs when the Green's function element $G_{ij}$ of the isolated system vanishes.
This condition leads to a complete suppression of the conductance ${\bf G}_{ij}$, regardless of the coupling parameters \( \tau_i \) and \( \tau_j \).}


\vspace{1em}
\section*{Acknowledgements}

We acknowledge financial support from the Core Program of the National Institute of Materials Physics,
granted by the Romanian MCID under projects
 no.\,PC2-PN23080202 and PC4-PN23080404.

\vspace{1em}


\footnotesize

\begin{thebibliography}{9}





\expandafter\ifx\csname url\endcsname\relax
  \def\url#1{\texttt{#1}}\fi
\expandafter\ifx\csname urlprefix\endcsname\relax\def\urlprefix{URL }\fi
\expandafter\ifx\csname href\endcsname\relax
  \def\href#1#2{#2} \def\path#1{#1}\fi

\bibitem{evers2020}
F.~Evers, R.~Koryt\'{a}r, S.~Tewari, J.~M. van Ruitenbeek, Advances and
  challenges in single-molecule electron transport, Rev. Mod. Phys.
  92 (2020) 035001.

\bibitem{tsuji2018}
Y.~Tsuji, E.~Estrada, R.~Movassagh, R.~Hoffmann, Quantum interference, graphs,
  walks, and polynomials, Chemical Reviews 118~(10) (2018) 4887--4911.

\bibitem{lambert2021}
C.~J. Lambert, {Quantum
  Transport in Nanostructures and Molecules}, 2053-2563, IOP Publishing, {(2021)}.

  
  
\bibitem{garner2020}
M.~H. Garner, G.~C. Solomon, Simultaneous suppression of $\pi$- and
  $\sigma$-transmission in $\pi$-conjugated molecules, The Journal of Physical
  Chemistry Letters 11~(17) (2020) 7400--7406.

\bibitem{gunasekaran2020}
S.~Gunasekaran, J.~E. Greenwald, L.~Venkataraman, Visualizing quantum
  interference in molecular junctions, Nano Lett. 20 (2020) 2843--2848.

\bibitem{yangli2019}
Y.~Li, X.~Yu, Y.~Zhen, H.~Dong, W.~Hu, Two-pathway viewpoint to interpret
  quantum interference in molecules containing five-membered heterocycles:
  Thienoacenes as examples, J. Phys. Chem. C 123 (2019) 15977--15984.

\bibitem{driscoll2021b}
L.~J. O'Driscoll, S.~Sangtarash, W.~Xu, A.~Daaoub, W.~Hong, H.~Sadeghi, M.~R.
  Bryce, Heteroatom effects on quantum interference in molecular junctions:
  Modulating antiresonances by molecular design, J. Phys. Chem. C 125 (2021)
  17385--17391.

\bibitem{pan2022}
H.~Pan, Y.~Wang, J.~Li, S.~Li, S.~Hou, Understanding quantum interference in
  molecular devices based on molecular conductance orbitals,
  {J. Phys. Chem. C}
 126~(40) (2022) 17424--17433.

 

\bibitem{valli2023}
A.~Valli, T.~Fabian, F.~Libisch, R.~Stadler,
{Stability
  of destructive quantum interference antiresonances in electron transport
  through graphene nanostructures}, Carbon 214 (2023) 118358.

  

\bibitem{lambert2015}
C.~J. Lambert, {Basic concepts of
  quantum interference and electron transport in single-molecule electronics},
  Chem. Soc. Rev. 44 (2015) 875--888.

  

\bibitem{kumar2024}
R.~Kumar, C.~Seth, R.~Venkatramani, V.~Kaliginedi,
  {Do quantum interference effects
  manifest in acyclic aliphatic molecules with anchoring groups?}, Nanoscale 15
  (2023) 15050--15058.

  

\bibitem{fan2024}
Y.~Fan, S.~Tao, S.~Pitié, C.~Liu, C.~Zhao, M.~Seydou, Y.~J. Dappe, P.~J. Low,
  R.~J. Nichols, L.~Yang,
  {Destructive quantum interference
  in meta-oligo(phenyleneethynylene) molecular wires with gold–graphene
  heterojunctions}, Nanoscale 16 (2024) 195--204.

  

\bibitem{qu2022}
F.-Y. Qu, Z.-H. Zhao, X.-R. Ren, S.-F. Zhang, L.~Wang, D.~Wang,
  {Multiple heteroatom substitution
  effect on destructive quantum interference in tripodal single-molecule
  junctions}, Phys. Chem. Chem. Phys. 24 (2022) 26795--26801.

  

\bibitem{ozlem2022}
O.~Sengul, J.~V\"olkle, A.~Valli, R.~Stadler,
 {Enhancing the
  sensitivity and selectivity of pyrene-based sensors for detection of small
  gaseous molecules via destructive quantum interference}, Phys. Rev. B 105
  (2022) 165428.

 

\bibitem{zhang2021}
B.~Zhang, M.~H. Garner, L.~Li, L.~M. Campos, G.~C. Solomon, L.~Venkataraman,
{Destructive quantum interference
  in heterocyclic alkanes: the search for ultra-short molecular insulators},
  Chem. Sci. 12 (2021) 10299--10305.

 
 
\bibitem{tada2002}
T.~Tada, K.~Yoshizawa, Quantum transport effects in nanosized graphite sheets,
  ChemPhysChem 3~(12) (2002) 1035--1037.

 

\bibitem{shuguang2018}
S.~Chen, G.~Chen, M.~A. Ratner,
  {Designing principles of
  molecular quantum interference effect transistors}, The Journal of Physical
  Chemistry Letters 9~(11) (2018) 2843--2847, pMID: 29750871.

\bibitem{cardamone2006}
D.~M. Cardamone, C.~A. Stafford, S.~Mazumdar, Controlling quantum transport
  through a single molecule, Nano Lett. 6 (2006) 2426.

\bibitem{bones2021}
D.~X. Bones, J.~T. Malme, E.~P. Hoy,
  Examining conductance values in the
 biphenyl molecular switch with reduced density matrices, {Int J Quantum Chem. 2021; 121:e26633.}



\bibitem{tianming2023}
T.~Li, V.~K. Bandari, O.~G. Schmidt,
  Molecular Electronics: Creating and Bridging Molecular Junctions and Promoting Its Commercialization,
  {Adv. Mater. 2023, 35, 2209088.}


\bibitem{alaa2023}
A.~A. Al-Jobory, A.~K. Ismael,
  {Controlling
  quantum interference in tetraphenyl-aza-bodipys}, Current Applied Physics 54
  (2023) 1--4.

 

\bibitem{chen2024}
Z.~Chen, I.~M. Grace, S.~L. Woltering, L.~Chen, A.~Gee, J.~Baugh, G.~A.~D.
  Briggs, L.~Bogani, J.~A. Mol, C.~J. Lambert, H.~L. Anderson, J.~O. Thomas,
  Quantum interference enhances the performance of single-molecule transistors,
  Nature Nanotechnology 19 (2024) 986--982.

\bibitem{hector2024}
H.~Vázquez, {Graphene edge
  interference improves single-molecule transistors}, Nature nanotechnology
  19~(7) (2024) 885—886.



\bibitem{fowler2009}
P.~W. Fowler, B.~T. Pickup, T.~Z. Todorova, W.~Myrvold, A selection rule for
  molecular conduction, {The Journal of Chemical Physics }131 (2009) 044104.

\bibitem{markussen2010}
T.~Markussen, R.~Stadler, K.~S. Thygesen,
 {The relation between structure and
  quantum interference in single molecule junctions}, Nano Letters 10~(10)
  (2010) 4260--4265, pMID: 20879779.


\bibitem{mayou2013}
D.~Mayou, Y.~Zhou, M.~Ernzerhof, The zero-voltage conductance of nanographenes:
  Simple rules and quantitative estimates, J. Phys. Chem. C 117 (2013) 7870.

\bibitem{stuyver2015}
T.~Stuyver, {S.~Fias}, F.~De~Proft, {P.~Geerlings}, Back of the envelope selection rule
  for molecular transmission: A curly arrow approach, J. Phys. Chem. C 119
  (2015) 26390.

\bibitem{nita2021}
M.~Ni\c{t}\u{a}, M.~\c{T}olea, D.~C. Marinescu, Conductance zeros in complex
  molecules and lattices from the interference set method, Phys. Rev. B 103
  (2021) 125307.

\bibitem{nita2022}
M.~Ni\c{t}\u{a}, D.~C. Marinescu, Persistent destructive quantum interference
  in the inverted graph method, Phys. Rev. B 105 (2022) 155303.

\bibitem{wallace1947}
P.~R. Wallace, {The band
  theory of graphite}, Phys. Rev. 71 (1947) 622--634.



\bibitem{alexander2013}
A.~Farrugia, J.~B. Gauci, I.~Sciriha,
{On the inverse of the adjacency
  matrix of a graph}, Special Matrices 1~(2013) (2013) 28--41 [cited
  2024-07-09].



\bibitem{baer2002}
R.~Baer, D.~Neuhauser, Anti-coherence based molecular electronics: Xor-gate
  response, Chemical Physics 281 (2002) 353.

\bibitem{tsuji2014}
Y.~Tsuji, R.~Hoffmann, R.~Movassagh, S.~Datta,
  {Quantum interference in polyenes},
  J Chem Phys. 141~(22) (2014) 224311.


\bibitem{pedersen2014}
K.~G.~L. Pedersen, M.~Strange, M.~Leijnse, P.~Hedeg\aa{}rd, G.~C. Solomon,
  J.~Paaske, {Quantum
  interference in off-resonant transport through single molecules}, Phys. Rev.
  B 90 (2014) 125413.



\bibitem{zhao2017}
X.~Zhao, V.~Geskin, R.~Stadler,
  {Destructive quantum interference in
  electron transport: A reconciliation of the molecular orbital and the atomic
  orbital perspective}, The Journal of Chemical Physics 146~(9) (2017) 092308.



\bibitem{junyang2018}
J.~Liu, X.~Huang, F.~Wang, W.~Hong,
  {Quantum interference
  effects in charge transport through single-molecule junctions: Detection,
  manipulation, and application}, Accounts of Chemical Research 52~(1) (2018)
  151--160.
  

\bibitem{garner2016}
M.~H. Garner, G.~C. Solomon, M.~Strange, Tunning conductance in aromatic
  molecules: Constructive and counteractive substituent effects, J Phys. Chem.
  C 120 (2016) 9097--9103.

\bibitem{sara2018}
S.~Sangtarash, H.~Sadeghi, C.~J. Lambert,
{Connectivity-driven
  bi-thermoelectricity in heteroatom-substituted molecular junctions}, Phys.
  Chem. Chem. Phys. 20 (2018) 9630--9637.

\bibitem{tsuji2019}
Y.~Tsuji, E.~Estrada, Influence of long-range interactions on quantum
  interference in molecular conduction. a tight-binding {(Hückel)} approach, The
  Journal of Chemical Physics 150 (2019) 204123.


\bibitem{ostahie2021}
B.~Ostahie, A.~Aldea,
 {Spectral
  analysis, chiral disorder and topological edge states manifestation in open
  non-hermitian su-schrieffer-heeger chains}, Physics Letters A 387 (2021)
  127030.


\bibitem{diventra2008}
Massimiliano~Di~Ventra, Electrical Transport in Nanoscale Systems,
Cambridge, Cambridge University Press (2008).







\bibitem{landauer1957}
R.~{Landauer}, Spatial variation of currents and fields due to localized
  scatterers in metallic conduction, IBM Journal of Research and Development
  1~(3) (1957) 223--231.


\bibitem{buttiker1986}
M.~B{\" u}ttiker, Four-terminal phase-coherent conductance, Phys. Rev. Lett 57
  (1986) 1761.

\bibitem{horia2005}
H.~D. Cornean, A.~Jensen, V.~Moldoveanu, {A rigorous proof of the
  Landauer–Büttiker formula}, Journal of Mathematical Physics 46~(4) (2005)
  042106.

\bibitem{tsuji2018b}
Y.~Tsuji, K.~Yoshizawa, Effects of electron-phonon coupling on quantum
  interference in polyenes, The Journal of Chemical Physics 149~(13) (2018)
  134115.


\bibitem{nita2018}
M.~Ni\c{t}\u{a}, B.~Ostahie, M.~\c{T}olea, A.~Aldea,
Localization Properties of Zig-Zag Edge States in Disordered Phosphorene,
physica status solidi (RRL) – Rapid Research Letters
12~ (7)~ 1800051 (2018).




\end{thebibliography}



\end{multicols}
\end{document}